\documentclass[useAMS,usenatbib]{mn2e}
\usepackage{graphicx}
\usepackage{amssymb}

\title[Source counts of SMGs at 1.1\,mm]{The source counts of submillimetre galaxies detected at $\lambda = 1.1$\,mm}
\author[Scott et al.]{K.S. Scott$^{1}$\thanks{E-mail:kscott@nrao.edu}, G.W. Wilson$^{2}$, I. Aretxaga$^3$, J.E. Austermann$^4$, E.L. Chapin$^5$, \newauthor J.S. Dunlop$^6$, H. Ezawa$^7$, M. Halpern$^5$, B. Hatsukade$^8$, D.H. Hughes$^3$, R. Kawabe$^9$, \newauthor S. Kim$^{10}$, K. Kohno$^{11,12}$, J.D. Lowenthal$^{13}$, A. Monta\~{n}a$^3$, K. Nakanishi$^{7,14,15}$, \newauthor T. Oshima$^9$, D. Sanders$^{16}$, D. Scott$^5$, N. Scoville$^{17}$, Y. Tamura$^{11}$, D. Welch$^2$, \newauthor M.S. Yun$^2$, M. Zeballos$^3$\\
$^{1}$North American ALMA Science Center, National Radio Astronomy Observatory, Charlottesville, VA 22903, USA \\
$^{2}$Department of Astronomy, University of Massachusetts, Amherst, MA 01003, USA \\
$^{3}$Instituto Nacional de Astrof\'{i}sica, \'{O}ptica y Electr\'{o}nica (INAOE), Aptdo. Postal 51 y 216, 72000 Puebla, Pue., Mexico \\
$^{4}$Center for Astrophysics and Space Astronomy, University of Colorado, Boulder, CO 80309, USA \\
$^{5}$Department of Physics \& Astronomy, University of British Columbia, 6224 Agricultural Road, Vancouver, BC V6T 1Z1, Canada \\
$^{6}$Institute for Astronomy, University of Edinburgh, Royal Observatory, Edinburgh EH9 3HJ, UK \\
$^{7}$ALMA Project Office, National Astronomical Observatory of Japan, 2-21-1 Osawa, Mitaka, Tokyo 181-8588, Japan \\
$^{8}$Department of Astronomy, Kyoto University, Kyoto 606-8502, Japan \\
$^{9}$Nobeyama Radio Observatory, National Astronomical Observatory of Japan, Minamimaki, Minamisaku, Nagano 384-1305, Japan \\
$^{10}$Astronomy and Space Science Department, Sejong University, Seoul, South Korea \\
$^{11}$Institute of Astronomy, University of Tokyo, 2-21-1 Osawa, Mitaka, Tokyo 181-0015, Japan \\
$^{12}$Research Center for the Early Universe, University of Tokyo, 7-3-1 Hongo, Bunkyo, Tokyo 113-0033, Japan \\
$^{13}$Department of Astronomy, Smith College, Northampton, MA 01063, USA \\
$^{14}$Joint ALMA Office, Alonso de Cordova 3107, Vitacura, Santiago 763 0355, Chile \\
$^{15}$The Graduate University for Advanced Studies (Sokendai), 2-21-1 Osawa, Mitaka, Tokyo 181-8588, Japan \\
$^{16}$Institute for Astronomy, 2680 Woodlawn Drive, University of Hawaii, Honolulu, HI 96822, USA \\
$^{17}$California Institute of Technology, MC 105-24, 1200 East California Boulevard, Pasadena, CA 91125, USA \\
}

\begin{document}

\date{}

\pagerange{\pageref{firstpage}--\pageref{lastpage}} \pubyear{2011}

\maketitle

\label{firstpage}

\begin{abstract}
The source counts of galaxies discovered at sub-millimetre and millimetre wavelengths provide important information on the evolution of infrared-bright galaxies. We combine the data from six blank-field surveys carried out at 1.1\,mm with AzTEC, totalling 1.6\,deg$^2$ in area with root-mean-square depths ranging from 0.4 to 1.7\,mJy, and derive the strongest constraints to date on the 1.1\,mm source counts at flux densities $S_{1100} = 1-12$\,mJy. Using additional data from the AzTEC Cluster Environment Survey to extend the counts to $S_{1100} \sim 20$\,mJy, we see tentative evidence for an enhancement relative to the exponential drop in the counts at $S_{1100} \sim 13$\,mJy and a smooth connection to the bright source counts at $>20$\,mJy measured by the South Pole Telescope; this excess may be due to strong lensing effects. We compare these counts to predictions from several semi-analytical and phenomenological models and find that for most the agreement is quite good at flux densities $\gtrsim 4$\,mJy; however, we find significant discrepancies ($\gtrsim3\sigma$) between the models and the observed 1.1\,mm counts at lower flux densities, and none of them are consistent with the observed turnover in the Euclidean-normalised counts at $S_{1100} \lesssim 2$\,mJy. Our new results therefore may require modifications to existing evolutionary models for low luminosity galaxies. Alternatively, the discrepancy between the measured counts at the faint end and predictions from phenomenological models could arise from limited knowledge of the spectral energy distributions of faint galaxies in the local Universe.
\end{abstract}


\section{Introduction}
\label{sec:intro}

Understanding how star formation evolved over the history of the Universe is one of the main goals of extragalactic astronomy today. Dust-obscured star formation is known to be a major contributor to the cosmic star formation history, as the cosmic infrared background (CIRB) accounts for $\sim50$\,per cent of the total extragalactic background light \citep{puget96}. Galaxies that are selected by their redshifted, thermal dust emission at sub-millimetre (submm) and millimetre (mm) wavelengths, hereafter SMGs \citep{smail97,hughes98,barger98}, are therefore thought to play a major role in the rapid build-up of the stellar populations within massive systems.

SMGs are predominantly high-redshift ($z \gtrsim 1$), dust-obscured galaxies whose far-infrared (FIR) luminosities ($L_{\mathrm{FIR}} \gtrsim 10^{12}\,\mathrm{L}_\odot$) imply high star formation rates (SFRs) of $\gtrsim 100\,\mathrm{M}_\odot$\,yr$^{-1}$; it is therefore generally believed that SMGs are observed during an important starburst or active phase in their evolution, en route to becoming massive elliptical galaxies at $z = 0$ \citep[see review by][]{blain02}. Since the rest-frame peak of the spectral energy distribution at $\lambda \sim 100\,\mu$m is increasingly redshifted into the submm/mm observing bands with increasing distance, there is a strong negative $k$-correction for surveys carried out at $\lambda \gtrsim 500\,\mu$m. SMGs over a wide range in redshift ($1 \lesssim z \lesssim 10$) are thus readily detected in deep, wide area surveys at these wavelengths, and consequently, their basic statistical properties -- such as their number density, redshift distribution, and clustering strength -- hold important clues to how the most massive galaxies assemble over time.

The source counts of SMGs as a function of flux density provide strong constraints for modelling the formation and evolution of IR-bright galaxies. There are two different methods for incorporating such constraints into models. The first method, often referred to as semi-analytical or forward evolution models, typically uses numerical simulations to describe the gravitational collapse of dark matter, combined with semi-analytical recipes to govern the evolution of baryonic processes within a galaxy, with some models including complex processes such as feedback from supernovae (SN) and/or active galactic nuclei (AGN) \citep[e.g.][]{granato04,baugh05,lacey10}. The second type are phenomenological models -- or parametric backward evolution models -- which make use of observational constraints (such as source counts and redshift distributions for galaxy populations selected at different wavelengths) to derive a model for the evolution of the luminosity function of galaxies, considering different populations of galaxies and spectral energy distributions (SEDs) \citep[e.g.][]{valiante09,rowanrobinson09,pearson09,bethermin11}. The source counts of SMGs measured from both long wavelength ($\lambda = 850-2000\,\mu$m) ground-based surveys \citep[e.g.][]{coppin06,bertoldi07,weiss09,austermann10,vieira10,marriage11} and shorter wavelength ($\lambda \le 500\,\mu$m) surveys from balloon- or space-based observatories \citep[e.g.][]{patanchon09,clements10,oliver10} require strong evolution in the properties of IR-bright galaxies.

In many ways, surveys at $24-500\,\mu$m from the {\it Spitzer Space Telescope} and the {\it Herschel Space Observatory} have surpassed those at longer wavelengths in terms of statistical power. The large areas combined with the significant depths of these surveys make them sensitive to a much broader range of galaxy types, whereas existing surveys at longer wavelengths are limited to only the most luminous systems. However, due to well known selection effects \citep[e.g.][]{blain02}, longer-wavelength data probe, on average, higher-redshift and/or colder galaxies, both of which are important components for understanding galaxy evolution. Several groups \citep[e.g.][]{devlin09,chary10} have shown that {\it Spitzer}/MIPS 24\,$\mu$m selected galaxies, which are predominately at $z \lesssim 1.5$, account for 55--95\,per cent of the CIRB at 70--500\,$\mu$m. However, these sources account for only $\sim30$\,per cent of the CIRB at $\lambda=1$\,mm \citep{scott10,penner11}. This suggests that galaxies selected at $24\,\mu$m, even in the deepest surveys, largely miss dust-obscured star formation activity at $z\gtrsim1.5$. The study of SMGs at $\lambda \ge 850\,\mu$m\ is thus essential to improving our understanding of the bulk of star formation taking place at higher redshifts.

There have been several deep (in some cases, confusion-limited) surveys carried out at 1.1\,mm with AzTEC on the James Clerk Maxwell Telescope (JCMT) and the Atacama Submillimeter Telescope Experiment \citep[ASTE;][]{ezawa04,ezawa08}. In this paper, we combine all previously published blank-field survey data taken with AzTEC in order to determine the strongest constraints to date on the number density of SMGs detected at $\lambda \ge 850\,\mu$m. Given the large total area of these combined fields, the uncertainty in the measured source counts from cosmic variance is very low. The 1.1\,mm source counts presented here thus provide important information on the highest-redshift, IR-bright galaxies, and can be used for improving models of galaxy evolution.

This paper is organised as follows: in Section~\ref{sec:fields}, we provide a summary of the blank-field surveys used to derive the combined-field source counts; in Section~\ref{sec:counts}, we describe the bootstrap sampling method used to derive the 1.1\,mm source counts, and we discuss how we incorporate systematic uncertainties from cosmic variance and flux calibration into our total error estimates; in Section~\ref{sec:brightcounts}, we discuss estimates of the source counts at very high flux densities determined from the South Pole Telescope (SPT) and other AzTEC surveys; we compare the combined-field 1.1\,mm source counts with predictions from current galaxy evolution models in Section~\ref{sec:galevol}; we discuss these results in Section~\ref{sec:disc}; and we summarise this work in Section~\ref{sec:conc}.

\section{Summary of AzTEC blank-field surveys}
\label{sec:fields}

We select the six individual blank-field surveys carried out with AzTEC \citep{wilson08} on the JCMT and ASTE from 2005 to 2008; each is briefly described below. Table~\ref{tab:summary} lists these fields, the telescope used, map area, depth, and number of SMGs detected in each. Note that we do not use AzTEC surveys of fields towards known over-densities, such as the AzTEC/ASTE map of the SSA-22 field towards a proto-cluster at $z = 3.1$ \citep{tamura09}, and the AzTEC/JCMT map of the MS-0451.6-0305 cluster at $z = 0.54$ \citep{wardlow10}. Our intent is to produce the strongest constraints on the {\it unbiased} 1.1\,mm source counts, extracted from ``blank" fields with no prior known over- or under-densities.

AzTEC map sensitivities tend to decrease from the map centre to the edges due to the scanning strategies typically employed. For uniformity, we consider the ``50\,per cent coverage region'' for all fields, which encompasses all pixels in a map for which the coverage (i.e. the inverse variance weight) is $\ge 50$\,per cent of the maximum coverage. This ensures that we are only considering regions of the maps that are well sampled by several detectors, with good cross-linking, and where the noise properties are uniform. The area and the range of $1\sigma$ root-mean-square (rms) depth for each field are listed in Table~\ref{tab:summary}. The combined fields result in a total area of 1.6\,deg$^2$ mapped to $1\sigma = 0.4$--1.7\,mJy\,beam$^{-1}$. The resolutions of the JCMT and ASTE data at $\lambda = 1.1$\,mm are $\theta = 18$\arcsec\ and 28\arcsec (full width at half maximum), respectively.

All of the AzTEC data were reduced using the standard customised data reduction pipeline in {\sc IDL}; this is described in detail in \citet{scott08}, so we do not describe it here. We recently derived an improvement in our estimated transfer function for point sources, as discussed in \citet{downes11}. This typically results in an increase of $10-30$\,per cent in the measured flux densities of point sources detected in the maps. This also increases the noise in the maps by roughly the same amount, such that the number of SMGs detected based on a signal-to-noise (S/N) threshold does not change significantly. \citet{downes11} provide revised source lists for the majority of previously published catalogues. This will also result in a shift of the source counts published prior to this correction, which affects all of the fields considered here with the exception of COSMOS \citep{aretxaga11}; however, the effect on the source counts is smaller, since the higher noise levels make flux boosting effects stronger \citep[see][]{austermann09,austermann10}, and in turn, the de-boosting corrections are larger.

\begin{table*}
\begin{center}
\caption{\label{tab:summary} Summary of AzTEC blank-field surveys\label{tab:summary}. The columns are: 1) the field name; 2) the telescope used for the survey; 3) the area of the survey within the 50\,per cent uniform coverage region (see Section~\ref{sec:fields}); 4) the range of rms noise within the 50\,per cent uniform coverage region; 5) the number of SMGs detected in the 50\,per cent uniform coverage region whose probability of de-boosting to $< 0$\,mJy is $P(S<0) \le 0.20$ for sources detected in JCMT surveys and $P(S<0) \le 0.05$ for sources detected in ASTE surveys (see Section~\ref{ssec:bootstrap}); 6) the root cosmic variance for the survey, following the estimate of \citet[][Section~\ref{ssec:cosmicvar}]{moster11}; and 7) the previous paper describing the survey.}
\begin{tabular}{llccccl}
\hline
Field & Telescope &{\it A}           & $\sigma_{\mathrm{rms}} $ & $N$                 & $\sigma_{\mathrm{gg}}$ & References \\
          &                     & (deg$^2$) & (mJy\,beam$^{-1}$)            &                          &                                              &                        \\

\hline
GOODS-N & JCMT & 0.08 & 1.2 -- 1.7 &  50 & 0.125 & \citet{perera08} \\
LH & JCMT & 0.30 & 1.1 -- 1.6 &  180 & 0.088 & \citet{austermann10} \\
GOODS-S & ASTE & 0.08 & 0.5 -- 0.8 &  66 & 0.129 & \citet{scott10} \\
ADF-S & ASTE & 0.20 & 0.4 -- 0.6 & 279 & 0.086 & \citet{hatsukade11} \\
SXDF & ASTE & 0.21 & 0.5 -- 0.7 & 271 & 0.098 & \citet{hatsukade11} \\
COSMOS & ASTE & 0.72 & 1.2 -- 1.7 & 230 & 0.065 & \citet{aretxaga11} \\
\hline
All &      & 1.60 & 0.4 -- 1.7 & 1076 & 0.039 &    \\
\hline
\end{tabular}
\end{center}
\end{table*}

\noindent {\it AzTEC/JCMT survey of the GOODS-N field} \hspace{3mm} The AzTEC survey of the Great Observatories Origins Deep North (GOODS-N) field was carried out during the 2005 to 2006 observing campaign on the JCMT and is presented in \citet{perera08}. GOODS-N is one of the most studied fields at all wavelengths, and much work has been done to identify the multi-wavelength counterparts to the SMGs discovered in the AzTEC survey \citep{chapin09}. These data have therefore been used extensively to characterise the redshift distribution, AGN fraction, etc. of mm-selected sources \citep[][Johnson et al. 2012]{yun11}. The revised catalogue for this survey, which incorporates the improved transfer function estimate, is presented in \citet{downes11}. 

\noindent {\it AzTEC/JCMT survey of the LH field} \hspace{3mm} A region in the Lockman Hole (LH) field was observed by AzTEC on the JCMT during the 2005 to 2006 observing campaign as part of the SCUBA HAlf-Degree Extragalactic Survey (SHADES) project and is described in \citet{austermann10}. SHADES consists of two discontiguous fields: the LH, and the Subaru/XMM-Newton Deep Field (SXDF). We do not use the AzTEC/JCMT map of the SXDF in our combined source counts analysis, since it largely overlaps with the ASTE survey of the same field (see below) and is considerably shallower. As with GOODS-N, the revised source catalogue for LH is presented in \citet{downes11}.

\noindent {\it AzTEC/ASTE survey of the GOODS-S field} \hspace{3mm} The AzTEC survey of the GOODS-South (GOODS-S) field was carried out on the ASTE telescope during the 2007 observing run and is presented in \citet{scott10}. As with GOODS-N, extensive efforts to identify multi-wavelength counterparts for AzTEC/GOODS-S SMGs, and to derive the redshift distribution, SFR, and stellar mass properties of SMGs from these data have already been carried out \citep[][Johnson et al. 2012]{yun11}. The revised catalogue using the new transfer function estimate is given in \citet{downes11}. In general, the fractional increase in the measured source flux densities and map noise is lower for the ASTE maps compared to the JCMT maps.

\noindent {\it AzTEC/ASTE survey of the ADF-S} \hspace{3mm} The AzTEC map of the {\it Akari} Deep Field South (ADF-S) was built up over the 2007 and 2008 observing runs on ASTE and is discussed in \citet{hatsukade11}. This is the deepest map used in our analysis and therefore puts strong constraints on the faint end of the source counts. 

\noindent {\it AzTEC/ASTE survey of the SXDF} \hspace{3mm} The AzTEC survey of the SXDF field carried out on ASTE during 2007 and 2008 is a slightly smaller but considerably deeper survey than the AzTEC SXDF map taken as part of the SHADES project on the JCMT. The source counts from this survey are presented in \citet{hatsukade11}, and, similar to the ADF-S, these data provide strong constraints on the faint end of the source counts. 

\noindent {\it AzTEC/ASTE survey of the COSMOS field} \hspace{3mm} The largest survey used in our combined source counts analysis is the AzTEC survey of the COSMOS field, carried out during the 2008 observing campaign on ASTE \citep{aretxaga11}. This survey almost completely encompasses the smaller, shallower AzTEC map taken with the JCMT in 2005 to 2006 \citep{scott08}, so we do not use the latter in our analysis. Being a factor of $>2$ larger than the other surveys considered here, this field provides the strongest constraints on the bright end of the 1.1\,mm source counts. The map used in our analysis here is the same as that presented in \citet{aretxaga11}, which used the improved transfer function of \citet{downes11}. Like the two GOODS fields, COSMOS is one of the best studied regions at all wavelengths, and the SMGs detected in this field have been used to study the properties of the SMG population at complementary wavelengths (Johnson et al. 2012).

\section{1.1\,mm source counts from combined blank-fields}
\label{sec:counts}

\subsection{Bootstrap sampling method}
\label{ssec:bootstrap}

To derive the 1.1\,mm source counts, we adopt the standard bootstrap sampling method that has been used extensively in the past for extracting the counts from AzTEC surveys. This method, first introduced by \citet{coppin06} and further developed for use with AzTEC data, is described in great detail in \citet{austermann09} and \citet{austermann10}, and we briefly summarise it here.

Using the source catalogue from one or more surveys and assuming a prior distribution for the source counts based on the best-fit Schechter function to the COSMOS counts \citep{aretxaga11}, we construct posterior flux distributions (PFDs) for each source that are sampled at random in order to determine intrinsic flux densities for the sources in the catalogue; these are then binned to derive the differential and integrated source counts. Only sources that pass the ``null threshold" test are sampled in order to avoid including a large number of false positives; that is, we only sample sources for which the probability that their intrinsic flux is less than zero is $P(S<0) \le 0.20$ for sources detected in JCMT surveys and $P(S<0) \le 0.05$ for sources detected in ASTE surveys. The more stringent limit for the ASTE data is imposed due to larger systematics from confusion in estimating the PFDs \citep[see discussion in][]{scott10}. This process is repeated 20,000 times in order to sufficiently sample the source count probability distribution, and the mean and 68.3\,per cent confidence interval for the counts in even-spaced, 1\,mJy-wide flux bins are computed from these iterations, giving the raw source counts. These raw counts are then corrected for incompleteness, which is estimated through simulation by calculating the recovery rate as a function of flux density for simulated sources injected (one at a time) into the map \citep[see][for example]{scott10}. We use the same source detection algorithm and null threshold test on these simulated sources as that used for the real catalogues to quantify the survey completeness. These corrected counts are then divided by the survey area to determine the differential ($dN/dS$) and integral ($N(>S)$) source counts. 

We compute the counts only for flux densities $\ge1$\,mJy; at lower flux densities, completeness is too low ($\lesssim10$\,per cent) and difficult to estimate. While the completeness in our three shallowest fields (GOODS-N, LH, and COSMOS) is $<5$\,per cent at 1\,mJy, the deeper surveys are 15--30\,per cent complete at this same flux level, ensuring that the low-flux end of the counts will not be subject to significant errors from biases in our completeness estimate.

Another common method for extracting source counts from this type of low-resolution, confusion-limited survey is the probability of deflection, or ``$P(D)$", approach, and it has been commonly employed for extracting counts from recent BLAST and {\it Herschel}-SPIRE surveys \citep[e.g.][]{patanchon09,glenn10}. The $P(D)$ technique avoids certain biases inherent in the bootstrap sampling method -- namely, the bias to the counts from the assumed prior distribution, and biases from the assumption that each detected ``source" really represents the emission from a single galaxy. Also, in principle, the $P(D)$ method allows an estimate of the source counts at fainter flux densities, below the detection limit of individual point sources. On the other hand, source counts determined from the $P(D)$ approach must use piecewise models, where the differential counts at selected ``nodes'' (i.e. fixed flux densities) are free parameters, and the nodes are connected by some smooth function. Such models may adequately reproduce the observed fluctuations in a map; however, they are not at all physically motivated. While increasing the resolution between nodes can reduce the model dependency of the counts, this increases the number of free parameters as well as the correlations between them. In practice, most groups limit the number of nodes so that the fitted parameters are largely uncorrelated, at the expense of making their results more model dependent; consequently, the formal errors on the fitted parameters may not always represent the true uncertainty in the counts \citep[see discussions in][for example]{scott10, glenn10}. Furthermore, while the implementation of the $P(D)$ method is relatively straightforward when the transfer function for point sources is linear, this is not the case for our PCA-cleaned AzTEC maps \citep{downes11}, and the $P(D)$ method thus becomes computationally expensive for our data. This is why we have chosen to use the bootstrap sampling method instead.

\citet{austermann10} demonstrated that for flux bins that are well sampled, the assumed prior used in the bootstrap sampling approach is quite weak. Furthermore, they show that biases to the counts for sparsely sampled flux bins can be effectively removed by an iterative process in which the counts extracted from the first pass are used to update the prior and PFDs for the source catalogue(s), and the bootstrap sampling method is repeated. We use this iterative process to decrease the effects of the prior on the extracted source counts. The counts from the combined blank fields change by less than 2\,per cent in all flux bins after only three iterations.

\begin{figure*}
\begin{center}
\includegraphics[width=13.5cm]{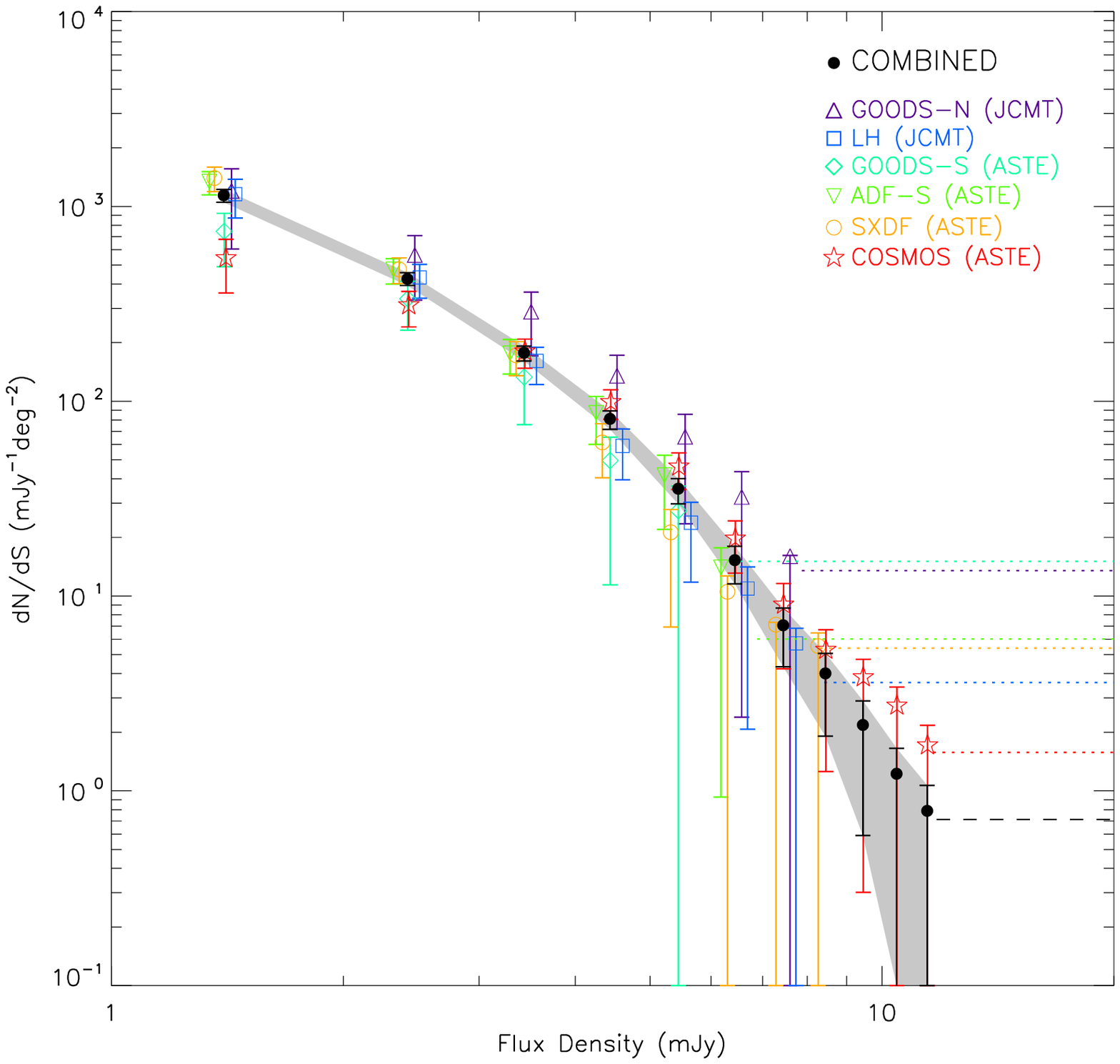}
\caption{Differential source counts derived from the six blank-field surveys carried out with AzTEC on the JCMT and ASTE. The counts determined from each individual field are as follows: {\it triangles} -- GOODS-N; {\it squares} -- LH; {\it diamonds} -- GOODS-S; {\it inverted triangles} -- ADF-S; {\it circles} -- SXDF; {\it stars} -- COSMOS. The counts derived from combining these six fields are shown as black filled circles. The counts for the individual blank-fields have been computed in slightly different flux bins for clarity in plotting. All error bars show the 68.3\,per cent confidence intervals determined from the bootstrap sampling method, including uncertainties arising from cosmic variance (see Section~\ref{ssec:cosmicvar}). The uncertainty from flux calibration (Section~\ref{ssec:calerror}) is not included since all fields were calibrated the same way. The shaded region highlights the 68.3\,per cent confidence range on the combined counts. The horizontal lines at the bottom right-hand side indicate the survey limits for each individual field (dotted) and for the combined counts (dashed). The survey limit corresponds to the expected value for which the source counts will Poisson deviate to zero  31.7\,per cent of the time, given the area of the survey(s).}
\label{fig:dnc_all}
\end{center}
\end{figure*}

We list in Table~\ref{tab:summary} the total number of sources in each field that were used to extract the counts. The counts from the combined blank-fields are shown in Fig.~\ref{fig:dnc_all} (filled circles), and are listed in Table~\ref{tab:allc}, where the upper and lower error bars indicate 68.3\,per cent confidence intervals. We give the correlation matrices \citep[see Appendix~A in][]{austermann10} for the differential and integrated source counts in Tables~\ref{tab:alldcorr} and \ref{tab:allicorr}, respectively, and we list the standard deviation on the counts, $\sigma_{dN/dS}$ and $\sigma_{N(>S)}$, for each bin as well, from which the covariance matrices can be determined. We have made the source counts and correlation matrices available online for public use.\footnote{http://www.astro.umass.edu/AzTEC/Scott2012\_nc/ \newline aztec\_combined\_counts\_2012.html} Since the counts are determined from bootstrapping off the PFDs of the sources, adjacent flux bins are strongly correlated. It is therefore important to use the covariance matrix in model fitting; e.g., for fitting a model prediction $\mathbf{m}$ to observed counts $\mathbf{d}$ with covariance matrix $\mathbf{C}$, the $\chi^2$ metric is given by $\chi^2 = (\mathbf{d}-\mathbf{m})\mathbf{C}^{-1}(\mathbf{d}-\mathbf{m})^{\mathrm{T}}$.

%%%Combined counts
\begin{table}
\begin{center}
\caption{1.1\,mm source counts\label{tab:allc} derived from the combined six AzTEC blank-field surveys. The first two columns show the flux bin centres and corresponding differential source counts, while the last two columns show the flux bin minima and cumulative counts. The first set of upper and lower errors shown on the counts indicate the 68.3\,per cent confidence intervals considering only statistical errors (Section~\ref{ssec:bootstrap}). The second set of errors in parentheses shows the 68.3\,per cent confidence intervals when including systematic uncertainties from cosmic variance (Section~\ref{ssec:cosmicvar}) and flux calibration (Section~\ref{ssec:calerror}). The bright source counts derived from the AzTEC Cluster Environment Survey (ACES) are also listed (see Section~\ref{sec:brightcounts}).}
\begin{tabular*}{0.5\textwidth}{@{\extracolsep{\fill}}cccc}
\hline
Flux Density  & $dN/dS$                              & Flux Density   & $N(>S)$        \\
(mJy)              & (mJy$^{-1}$\,deg$^{-2}$) & (mJy)                &  (deg$^{-2}$) \\
\hline
\multicolumn{4}{c}{Combined blank-fields} \\
\hline
$ 1.4$ & $1140^{+70}_{-80}(^{+100}_{-120})$ & $ 1.0$ & $1890^{+70}_{-70}(^{+110}_{-120})$ \\  [+3pt]
$ 2.4$ & $420^{+30}_{-30}(^{+30}_{-30})$ & $ 2.0$ & $750^{+30}_{-30}(^{+50}_{-50})$ \\  [+3pt]
$ 3.4$ & $180^{+10}_{-10}(^{+20}_{-20})$ & $ 3.0$ & $330^{+20}_{-20}(^{+30}_{-30})$ \\  [+3pt]
$ 4.4$ & $81^{+8}_{-8}(^{+9}_{-11})$ & $ 4.0$ & $150^{+10}_{-10}(^{+20}_{-20})$ \\  [+3pt]
$ 5.4$ & $36^{+5}_{-5}(^{+7}_{-7})$ & $ 5.0$ & $67^{+6}_{-6}(^{+12}_{-13})$ \\  [+3pt]
$ 6.4$ & $15^{+3}_{-3}(^{+3}_{-5})$ & $ 6.0$ & $32^{+4}_{-4}(^{+6}_{-8})$ \\  [+3pt]
$ 7.4$ & $7^{+2}_{-2}(^{+2}_{-3})$ & $ 7.0$ & $17^{+3}_{-3}(^{+4}_{-4})$ \\  [+3pt]
$ 8.4$ & $4.0^{+1.4}_{-1.8}(^{+1.1}_{-2.3})$ & $ 8.0$ & $9^{+2}_{-2}(^{+2}_{-3})$ \\  [+3pt]
$ 9.4$ & $2.2^{+0.9}_{-1.3}(^{+0.6}_{-1.8})$ & $ 9.0$ & $5.5^{+1.3}_{-1.8}(^{+1.7}_{-2.1})$ \\  [+3pt]
$10.4$ & $1.2^{+0.6}_{-1.0}(^{+0.4}_{-1.2})$ & $10.0$ & $3.3^{+1.2}_{-1.5}(^{+1.4}_{-1.6})$ \\  [+3pt]
$11.4$ & $0.8^{+0.4}_{-0.8}(^{+0.3}_{-0.8})$ & $11.0$ & $2.1^{+0.3}_{-2.1}(^{+1.1}_{-1.8})$ \\  [+3pt]
\hline
\multicolumn{4}{c}{ACES fields} \\
\hline
$11.1$ & $0.6^{+0.2}_{-0.3}(^{+0.2}_{-0.3})$ & $10.0$ & $3.0^{+0.9}_{-0.7}(^{+1.0}_{-1.0})$ \\   [+3pt]
$14.1$ & $0.3^{+0.1}_{-0.2}(^{+0.1}_{-0.2})$ & $13.0$ & $1.28^{+0.02}_{-1.22}(^{+0.03}_{-1.21})$ \\   [+3pt]
$17.1$ & $0.15^{+0.07}_{-0.15}(^{+0.07}_{-0.15})$ & $16.0$ & $0.51^{+0.01}_{-0.45}(^{+0.03}_{-0.51})$ \\  [+3pt]
\hline
\end{tabular*}
\end{center}
\end{table}

\begin{table*}
\begin{center}
\caption{Correlation matrix for the differential counts\label{tab:alldcorr} derived from the combined AzTEC blank-field surveys. The first set shows the bin-to-bin correlations when considering only statistical errors from the bootstrap sampling method (Section~\ref{ssec:bootstrap}); these represent the actual correlations between bins in the data themselves. The second, third, and fourth sets show the correlations when systematic uncertainties from cosmic variance (Section~\ref{ssec:cosmicvar}), flux calibration (Section~\ref{ssec:calerror}), and both are included, respectively. The last column in all four cases shows the standard deviation on the differential counts for each flux bin, and can be used to compute the covariance matrix from these correlations.}
%{ \scriptsize
\begin{tabular}{lcccccccccccc}
\hline
Flux Density  &           &           &           &           &           &           &           &           &           &            &          & $\sigma_{dN/dS}$ \\
(mJy) &  1.4 &  2.4 &  3.4 &  4.4 &  5.4 &  6.4 &  7.4 &  8.4 &  9.4 & 10.4 & 11.4 & (mJy$^{-1}$\,deg$^{-2}$) \\
\hline
  & \multicolumn{12}{l}{Statistical errors only} \\ 
\hline
 1.4 &  1.00 &     &     &     &     &     &     &     &     &     &     & 66 \\
 2.4 &  0.61 &  1.00 &     &     &     &     &     &     &     &     &     & 27 \\
 3.4 &  0.21 &  0.62 &  1.00 &     &     &     &     &     &     &     &     & 14 \\
 4.4 &  0.11 &  0.29 &  0.67 &  1.00 &     &     &     &     &     &     &     & 8.2 \\
 5.4 &  0.06 &  0.15 &  0.34 &  0.71 &  1.00 &     &     &     &     &     &     & 5.0 \\
 6.4 &  0.02 &  0.06 &  0.15 &  0.33 &  0.67 &  1.00 &     &     &     &     &     & 3.2 \\
 7.4 &  0.00 &  0.01 &  0.05 &  0.13 &  0.32 &  0.72 &  1.00 &     &     &     &     & 2.1 \\
 8.4 & -0.01 & -0.01 &  0.00 &  0.04 &  0.12 &  0.32 &  0.76 &  1.00 &     &     &     & 1.6 \\
 9.4 &  0.00 & -0.01 & -0.01 &  0.01 &  0.05 &  0.17 &  0.49 &  0.81 &  1.00 &     &     & 1.2 \\
10.4 &  0.00 & -0.01 & -0.02 & -0.01 &  0.01 &  0.06 &  0.20 &  0.42 &  0.83 &  1.00 &     & 0.87 \\
11.4 &  0.00 & -0.01 & -0.02 & -0.01 & -0.01 &  0.01 &  0.07 &  0.20 &  0.54 &  0.85 &  1.00 & 0.70 \\
\hline
 & \multicolumn{12}{l}{Including systematic uncertainties from cosmic variance} \\ 
\hline
 1.4 &  1.00 &     &     &     &     &     &     &     &     &     &     & 80 \\
 2.4 &  0.80 &  1.00 &     &     &     &     &     &     &     &     &     & 31 \\
 3.4 &  0.57 &  0.79 &  1.00 &     &     &     &     &     &     &     &     & 15 \\
 4.4 &  0.48 &  0.58 &  0.79 &  1.00 &     &     &     &     &     &     &     & 8.8 \\
 5.4 &  0.38 &  0.44 &  0.54 &  0.79 &  1.00 &     &     &     &     &     &     & 5.2 \\
 6.4 &  0.27 &  0.30 &  0.35 &  0.47 &  0.72 &  1.00 &     &     &     &     &     & 3.2 \\
 7.4 &  0.19 &  0.21 &  0.22 &  0.27 &  0.40 &  0.73 &  1.00 &     &     &     &     & 2.1 \\
 8.4 &  0.13 &  0.14 &  0.14 &  0.14 &  0.19 &  0.36 &  0.77 &  1.00 &     &     &     & 1.6 \\
 9.4 &  0.11 &  0.12 &  0.11 &  0.11 &  0.12 &  0.22 &  0.51 &  0.81 &  1.00 &     &     & 1.2 \\
10.4 &  0.09 &  0.09 &  0.08 &  0.07 &  0.07 &  0.09 &  0.22 &  0.43 &  0.83 &  1.00 &     & 0.87 \\
11.4 &  0.07 &  0.08 &  0.07 &  0.06 &  0.05 &  0.05 &  0.09 &  0.22 &  0.55 &  0.85 &  1.00 & 0.71 \\
\hline
 & \multicolumn{12}{l}{Including systematic uncertainties from flux calibration} \\ 
\hline
 1.4 &  1.00 &     &     &     &     &     &     &     &     &     &     & 94 \\
 2.4 &  0.58 &  1.00 &     &     &     &     &     &     &     &     &     & 27 \\
 3.4 &  0.20 &  0.62 &  1.00 &     &     &     &     &     &     &     &     & 14 \\
 4.4 &  0.11 &  0.27 &  0.67 &  1.00 &     &     &     &     &     &     &     & 9.3 \\
 5.4 &  0.05 &  0.14 &  0.33 &  0.71 &  1.00 &     &     &     &     &     &     & 6.5 \\
 6.4 &  0.02 &  0.06 &  0.16 &  0.32 &  0.68 &  1.00 &     &     &     &     &     & 4.0 \\
 7.4 &  0.00 &  0.02 &  0.06 &  0.13 &  0.30 &  0.67 &  1.00 &     &     &     &     & 2.5 \\
 8.4 &  0.00 &  0.00 &  0.01 &  0.03 &  0.11 &  0.34 &  0.83 &  1.00 &     &     &     & 1.7 \\
 9.4 & -0.01 &  0.00 &  0.00 &  0.01 &  0.04 &  0.16 &  0.46 &  0.75 &  1.00 &     &     & 1.3 \\
10.4 &  0.00 &  0.01 &  0.00 &  0.00 &  0.00 &  0.04 &  0.17 &  0.42 &  0.86 &  1.00 &     & 0.92 \\
11.4 &  0.01 &  0.01 &  0.01 &  0.00 &  0.00 &  0.00 &  0.05 &  0.17 &  0.48 &  0.79 &  1.00 & 0.72 \\
\hline
 & \multicolumn{12}{l}{Including systematic uncertainties from cosmic variance and flux calibration} \\ 
\hline
 1.4 &  1.00 &     &     &     &     &     &     &     &     &     &     & 100 \\
 2.4 &  0.78 &  1.00 &     &     &     &     &     &     &     &     &     & 32 \\
 3.4 &  0.57 &  0.80 &  1.00 &     &     &     &     &     &     &     &     & 16 \\
 4.4 &  0.47 &  0.58 &  0.79 &  1.00 &     &     &     &     &     &     &     & 9.9 \\
 5.4 &  0.38 &  0.44 &  0.55 &  0.79 &  1.00 &     &     &     &     &     &     & 6.7 \\
 6.4 &  0.26 &  0.30 &  0.34 &  0.45 &  0.73 &  1.00 &     &     &     &     &     & 4.0 \\
 7.4 &  0.18 &  0.19 &  0.20 &  0.24 &  0.38 &  0.69 &  1.00 &     &     &     &     & 2.5 \\
 8.4 &  0.13 &  0.14 &  0.14 &  0.15 &  0.21 &  0.40 &  0.85 &  1.00 &     &     &     & 1.7 \\
 9.4 &  0.12 &  0.12 &  0.12 &  0.12 &  0.14 &  0.22 &  0.46 &  0.75 &  1.00 &     &     & 1.3 \\
10.4 &  0.09 &  0.10 &  0.10 &  0.09 &  0.09 &  0.11 &  0.20 &  0.46 &  0.86 &  1.00 &     & 0.91 \\
11.4 &  0.08 &  0.09 &  0.08 &  0.07 &  0.07 &  0.06 &  0.07 &  0.18 &  0.44 &  0.76 &  1.00 & 0.72 \\
\hline
\end{tabular}
%}
\end{center}
\end{table*}

\begin{table*}
\begin{center}
\caption{Correlation matrix for the cumulative counts\label{tab:allicorr} derived from the combined AzTEC blank-field surveys. The first set shows the bin-to-bin correlations when considering only statistical errors from the bootstrap sampling method (Section~\ref{ssec:bootstrap}); these represent the actual correlations between bins in the data themselves. The second, third, and fourth sets show the correlations when systematic uncertainties from cosmic variance (Section~\ref{ssec:cosmicvar}), flux calibration (Section~\ref{ssec:calerror}), and both are included, respectively. The last column in all four cases shows the standard deviation on the cumulative counts for each flux bin, and can be used to compute the covariance matrix from these correlations.}
%{ \scriptsize
\begin{tabular}{lcccccccccccc}
\hline
Flux Density  &           &           &           &           &           &           &           &           &           &            &         & $\sigma_{N(>S)}$              \\
(mJy) &  1.0 &  2.0 &  3.0 &  4.0 &  5.0 &  6.0 &  7.0 &  8.0 &  9.0 & 10.0 & 11.0 & (deg$^{-2}$) \\
\hline
  & \multicolumn{12}{l}{Statistical errors only} \\ 
\hline
 1.0 &  1.00 &     &     &     &     &     &     &     &     &     &     & 74 \\
 2.0 &  0.77 &  1.00 &     &     &     &     &     &     &     &     &     & 32 \\
 3.0 &  0.47 &  0.83 &  1.00 &     &     &     &     &     &     &     &     & 17 \\
 4.0 &  0.31 &  0.59 &  0.87 &  1.00 &     &     &     &     &     &     &     & 11 \\
 5.0 &  0.19 &  0.40 &  0.64 &  0.88 &  1.00 &     &     &     &     &     &     & 6.7 \\
 6.0 &  0.11 &  0.24 &  0.43 &  0.66 &  0.90 &  1.00 &     &     &     &     &     & 4.5 \\
 7.0 &  0.06 &  0.15 &  0.28 &  0.48 &  0.72 &  0.92 &  1.00 &     &     &     &     & 3.2 \\
 8.0 &  0.04 &  0.10 &  0.21 &  0.36 &  0.57 &  0.78 &  0.94 &  1.00 &     &     &     & 2.4 \\
 9.0 &  0.03 &  0.07 &  0.15 &  0.27 &  0.44 &  0.63 &  0.80 &  0.93 &  1.00 &     &     & 1.8 \\
10.0 &  0.02 &  0.05 &  0.11 &  0.21 &  0.34 &  0.49 &  0.65 &  0.80 &  0.95 &  1.00 &     & 1.4 \\
11.0 &  0.02 &  0.04 &  0.08 &  0.16 &  0.26 &  0.38 &  0.52 &  0.66 &  0.84 &  0.95 &  1.00 & 1.1 \\
\hline
  & \multicolumn{12}{l}{Including systematic uncertainties from cosmic variance} \\ 
\hline
 1.0 &  1.00 &     &     &     &     &     &     &     &     &     &     & 100 \\
 2.0 &  0.90 &  1.00 &     &     &     &     &     &     &     &     &     & 43 \\
 3.0 &  0.74 &  0.91 &  1.00 &     &     &     &     &     &     &     &     & 21 \\
 4.0 &  0.61 &  0.76 &  0.91 &  1.00 &     &     &     &     &     &     &     & 12 \\
 5.0 &  0.46 &  0.58 &  0.73 &  0.90 &  1.00 &     &     &     &     &     &     & 7.2 \\
 6.0 &  0.32 &  0.41 &  0.53 &  0.70 &  0.90 &  1.00 &     &     &     &     &     & 4.7 \\
 7.0 &  0.23 &  0.29 &  0.38 &  0.52 &  0.73 &  0.92 &  1.00 &     &     &     &     & 3.3 \\
 8.0 &  0.17 &  0.21 &  0.28 &  0.40 &  0.58 &  0.79 &  0.95 &  1.00 &     &     &     & 2.5 \\
 9.0 &  0.13 &  0.16 &  0.21 &  0.30 &  0.45 &  0.63 &  0.80 &  0.93 &  1.00 &     &     & 1.9 \\
10.0 &  0.10 &  0.12 &  0.16 &  0.23 &  0.35 &  0.49 &  0.65 &  0.80 &  0.95 &  1.00 &     & 1.4 \\
11.0 &  0.07 &  0.09 &  0.13 &  0.18 &  0.27 &  0.39 &  0.51 &  0.66 &  0.83 &  0.95 &  1.00 & 1.1 \\
\hline
  & \multicolumn{12}{l}{Including systematic uncertainties from flux calibration} \\ 
\hline
 1.0 &  1.00 &     &     &     &     &     &     &     &     &     &     & 90 \\
 2.0 &  0.75 &  1.00 &     &     &     &     &     &     &     &     &     & 35 \\
 3.0 &  0.46 &  0.83 &  1.00 &     &     &     &     &     &     &     &     & 24 \\
 4.0 &  0.30 &  0.59 &  0.87 &  1.00 &     &     &     &     &     &     &     & 18 \\
 5.0 &  0.19 &  0.39 &  0.63 &  0.88 &  1.00 &     &     &     &     &     &     & 12 \\
 6.0 &  0.11 &  0.25 &  0.43 &  0.65 &  0.89 &  1.00 &     &     &     &     &     & 7.1 \\
 7.0 &  0.07 &  0.16 &  0.29 &  0.47 &  0.71 &  0.92 &  1.00 &     &     &     &     & 4.5 \\
 8.0 &  0.05 &  0.11 &  0.21 &  0.36 &  0.57 &  0.78 &  0.94 &  1.00 &     &     &     & 3.1 \\
 9.0 &  0.03 &  0.08 &  0.16 &  0.27 &  0.42 &  0.60 &  0.78 &  0.93 &  1.00 &     &     & 2.2 \\
10.0 &  0.03 &  0.07 &  0.12 &  0.20 &  0.32 &  0.47 &  0.63 &  0.80 &  0.95 &  1.00 &     & 1.6 \\
11.0 &  0.02 &  0.05 &  0.10 &  0.16 &  0.25 &  0.37 &  0.50 &  0.66 &  0.83 &  0.95 &  1.00 & 1.2 \\
\hline
  & \multicolumn{12}{l}{Including systematic uncertainties from cosmic variance and flux calibration} \\ 
\hline
 1.0 &  1.00 &     &     &     &     &     &     &     &     &     &     & 120 \\
 2.0 &  0.89 &  1.00 &     &     &     &     &     &     &     &     &     & 45 \\
 3.0 &  0.74 &  0.91 &  1.00 &     &     &     &     &     &     &     &     & 27 \\
 4.0 &  0.60 &  0.75 &  0.91 &  1.00 &     &     &     &     &     &     &     & 19 \\
 5.0 &  0.46 &  0.57 &  0.72 &  0.90 &  1.00 &     &     &     &     &     &     & 12 \\
 6.0 &  0.32 &  0.40 &  0.52 &  0.69 &  0.90 &  1.00 &     &     &     &     &     & 7.2 \\
 7.0 &  0.23 &  0.28 &  0.37 &  0.52 &  0.73 &  0.92 &  1.00 &     &     &     &     & 4.5 \\
 8.0 &  0.18 &  0.23 &  0.29 &  0.41 &  0.59 &  0.79 &  0.94 &  1.00 &     &     &     & 3.1 \\
 9.0 &  0.14 &  0.18 &  0.23 &  0.31 &  0.46 &  0.62 &  0.78 &  0.93 &  1.00 &     &     & 2.2 \\
10.0 &  0.12 &  0.14 &  0.18 &  0.24 &  0.35 &  0.48 &  0.63 &  0.81 &  0.95 &  1.00 &     & 1.6 \\
11.0 &  0.09 &  0.12 &  0.14 &  0.19 &  0.28 &  0.38 &  0.50 &  0.66 &  0.83 &  0.96 &  1.00 & 1.3 \\
\hline
\end{tabular}
%}
\end{center}
\end{table*}

\subsection{Effects of confusion on the extracted source counts}
\label{ssec:confusion}

Using the standard definition of one source per 30 beams \citep[e.g.][]{takeuchi04}, the confusion limit for these surveys is $S_{\mathrm{lim}} = $\ 1.4 and 2.4\,mJy for those carried out on the JCMT and ASTE, respectively. These correspond to the two lowest flux bins in our source counts estimate. In this section, we explore potential biases to the extracted counts due to confusion effects.

For this purpose, we make fully simulated data-sets for each of the six blank fields. These simulated maps have the same noise properties as the real data, and are all populated with the same source distribution described by a Schechter function:

\begin{equation}
{dN \over dS} = N_{3\mathrm{mJy}} \left({S \over 3\,\mathrm{mJy}}\right)^{\alpha+1} e^{ -(S - 3\,\mathrm{mJy})/S^\prime }, 
\label{eqn:schechter}
\end{equation}

\noindent We use values of $N_{3\mathrm{mJy}} = 230$\,mJy$^{-1}$\,deg$^{-2}$, $S^\prime = 1.7$\,mJy, and $\alpha = -2$, which provide a good fit to the observed blank-field source counts derived in this paper. Since we are only looking to examine general trends in potential biases to the counts from source confusion, {\it a priori} knowledge of the true distribution of the underlying counts is not necessary. We populate each simulated map with sources down to a flux density limit of 0.1\,mJy, where the cumulative counts reach $>1$ source per beam for both JCMT and ASTE surveys. Previous studies \citep[e.g.][]{scott10} have demonstrated that the choice of this lower flux density cutoff is not too critical, so long as it corresponds to where the integrated source counts are $\gtrsim1$ source per beam, since adding fainter sources at that point would not change the flux distribution in the map. Simulated sources are placed at random positions drawn from a uniform spatial distribution. We make 100 simulated maps for each field and use the bootstrap sampling method described in Section~\ref{ssec:bootstrap} to derive the source counts for each of them.

\begin{figure}
\begin{center}
\includegraphics[width=8.5cm]{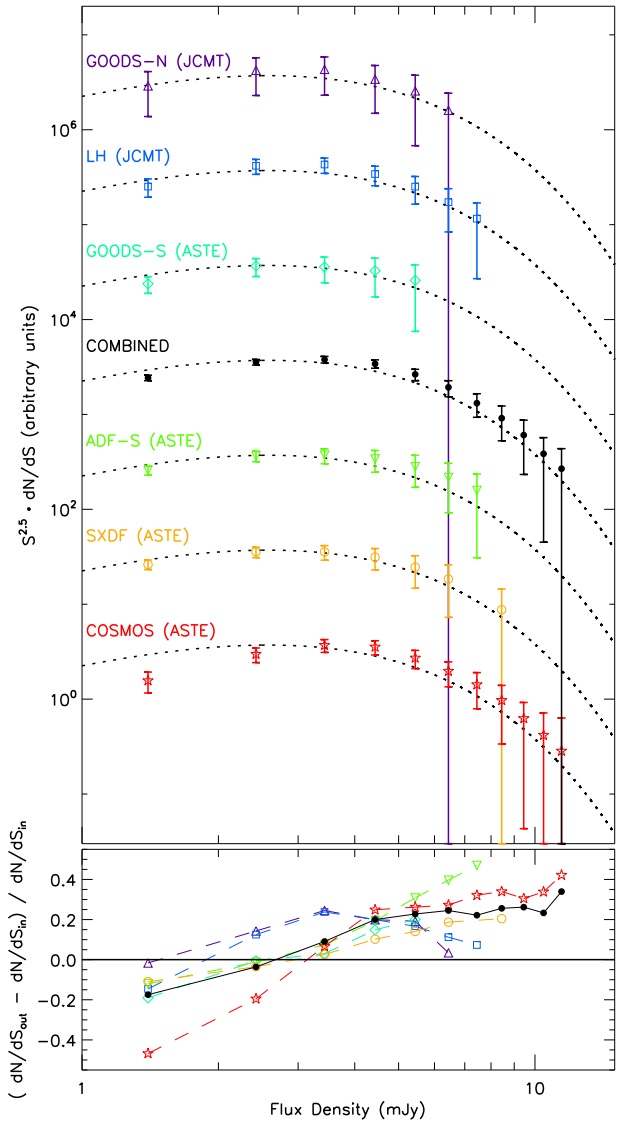}
\caption{Results of simulations to test the effects of confusion on our source counts extraction, as described in Section~\ref{ssec:confusion}. The top panel shows the averaged Euclidean-normalised differential source counts extracted from simulated maps for each field and for the combined fields, as labeled. For clarity, the counts have been offset by a factor of 1000, 100, 10, 0.1, 0.01, and 0.001 for the GOODS-N, LH, GOODS-S, ADF-S, SXDF, and COSMOS data-sets, respectively. The dotted curves show the model for the input source counts (Equation~\ref{eqn:schechter}) for comparison. The errors represent typical 68.3\,per cent confidence intervals on the counts {\it for a single simulated map}, demonstrating that the biases to the extracted counts arising from confusion effects are small compared to the statistical errors. The bottom panel shows the fractional difference between the input and output source counts, using the same symbols as in the top panel.}
\label{fig:sims}
\end{center}
\end{figure}

The results from these simulations are presented in Fig.~\ref{fig:sims}, which shows the Euclidean-normalised differential source counts averaged over the 100 simulated maps for each field separately (top panel). The counts have been scaled arbitrarily for clarity, with dotted curves indicating the input source distribution from Equation~\ref{eqn:schechter}. There is evidence from these simulations that confusion introduces biases to the observed source counts. This is more evident in the bottom panel of Fig.~\ref{fig:sims}, which shows the fractional difference between the input and output source counts. In general, the counts at $S_{1100} \gtrsim S_{\mathrm{lim}}$ (the confusion limit) are overestimated by $\sim5$-30\,per cent, while the counts at $S_{1100} \lesssim S_{\mathrm{lim}}$ are underestimated by $\sim10$-20\,per cent. However, these biases are small compared to the statistical errors on the derived counts for these individual fields; the error bars in the top panel of Fig.~\ref{fig:sims} represent the typical 68.3\,per cent confidence intervals on the extracted counts {\it for a single simulated map}. Considering all simulated data-sets, the extracted counts agree with the input counts within their $2\sigma$ errors $>85$\,per cent of the time, with the exception of the COSMOS simulated fields, where the extracted counts at $S_{1100} = 1.4$\,mJy are always significantly underestimated. This large discrepancy between the input and output source counts in the lowest flux density bin for COSMOS is most likely due to the low -- and therefore poorly measured -- completeness at that flux density.

We next use the simulated maps for each field to make 100 realisations of the extracted source counts from the six fields combined. For each realisation, we randomly select six simulated maps, one from each field, and carry out the joint bootstrap sampling extraction. These results are also shown in Fig.~\ref{fig:sims}, and as expected, we see similar biases to the output source counts as seen in each individual field. This bias is smallest (4\,per cent) for the 2.4\,mJy flux density bin, which corresponds to the confusion limit for the ASTE surveys. For the lowest flux bin at 1.4\,mJy, the counts are underestimated by 17\,per cent, and for the bins at $S_{1100} > 3$\,mJy, the counts are overestimated by 9-34\,per cent. In comparison to the statistical errors, the combined-field counts derived from the simulated maps agree with the input counts within their $2\sigma$ errors $>80$\,per cent of the time.

\subsection{Cosmic variance}
\label{ssec:cosmicvar}

The source counts determined from each individual survey are shown in Fig.~\ref{fig:dnc_all}. Given the limited area surveyed for each field, we expect to see variations in the counts from field-to-field owing to variations in the underlying large-scale structure (also known as ``cosmic variance"). Furthermore, given the strong bin-to-bin correlations, the counts across all flux bins for a given survey vary in the same sense; for example, the GOODS-N counts are consistently higher than the average, while the GOODS-S counts are consistently lower. In order to assess the agreement among the counts derived from individual AzTEC fields, we must include this uncertainty from cosmic variance into the error budget. 

We estimate the expected cosmic variance for each individual survey and for the combined blank-fields using the prescription described in \citet{moster11}. This estimate uses predictions of the underlying structure of cold dark matter and the expected bias for a galaxy population -- in this case, SMGs. This simple recipe depends only on the angular dimensions of the field ($\alpha_1$, $\alpha_2$), the mean redshift ($\bar{z}$), redshift bin size ($\Delta z$), and stellar mass ($M_\star$) of the galaxy population in question. \citet{moster11} have provided their software tools for calculating cosmic variance online.\footnote{http://www.mpia.de/homes/moster/research}

This estimate assumes rectangular geometry for the survey, which is rarely the case for our fields; however, \citet{moster11} show that the geometry makes little difference except where the ratio between the short and long axes of the survey is $\lesssim 0.2$, which is not the case for any of our fields. We therefore assume angular dimensions for each field, $\alpha_1$ and $\alpha_2$, such that the product is equal to the area in the 50\,per cent uniform coverage region, and the ratio approximately matches the geometry of the AzTEC map. The mean redshift, redshift bin size, and stellar mass are taken from \citet{yun11}, who use spectroscopic and photometric redshifts for the SMGs detected in both AzTEC/GOODS fields to determine their redshift distribution, and stellar mass estimates from modelling their observed rest-frame UV and optical SEDs. \citet{yun11} find $\bar{z} = 2.5$, $\Delta z = 1.5$, and $M_\star > 10^{10}\,\mathrm{M}_\odot$, where $\Delta z$ and the limit on $M_\star$ encompass 75\,per cent of the SMGs in that sample. The root cosmic variance, $\sigma_{\mathrm{gg}}$, which represents the expected fractional error on the counts due to cosmic variance, is listed for each field in Table~\ref{tab:summary} and ranges from 6.5--12.9\,per cent for the individual AzTEC fields. By combining all six fields, totalling 1.6\,deg$^2$, uncertainties due to cosmic variance are reduced to only 3.9\,per cent, which is smaller than the statistical errors on the counts ($\ge6$\,per cent).

Since the uncertainty from cosmic variance is completely correlated among all flux bins, we cannot simply add $\sigma_{\mathrm{gg}}$ in quadrature with the statistical errors on the counts; instead, we must consider how including cosmic variance affects the entire covariance matrix. It is straightforward to include this effect within the framework of the bootstrap sampling method. For {\it each} of the 20,000 iterations, we generate a random number drawn from a Gaussian distribution with a mean of zero and a standard deviation of $\sigma_{\mathrm{gg}}$, and we apply this fractional deviation to the differential source counts {\it uniformly to all bins}. The mean, 68.3\,per cent confidence intervals, and the covariance matrix for the counts are then computed from the 20,000 iterations in the same manner as in the standard bootstrap method described in Section~\ref{ssec:bootstrap}. This way, we broaden the distribution in the bootstrapped counts according to the expected degree of cosmic variance and can properly trace the effects on the bin-to-bin correlations.

The 68.3\,per cent confidence intervals shown by the error bars on the differential counts in Fig.~\ref{fig:dnc_all} include the uncertainties expected from cosmic variance, for each individual field as well as the combined fields. We note that the standard deviation of the counts, $\sigma_{dN/dS}$ (equal to the root of the diagonal elements of the covariance matrix), increases as expected ($\sigma^2_{dN/dS} \rightarrow \sigma^2_{dN/dS} + \sigma^2_{\mathrm{gg}}(dN/dS)^2$), and the bin-to-bin correlations on the counts increase substantially, as shown in Tables~\ref{tab:alldcorr} and \ref{tab:allicorr}. Comparing the counts observed in each individual field when both the statistical errors and the uncertainties from cosmic variance are included, we find that they all agree quite well. 

The mean redshift and inter-quartile range from \citet{yun11} which we use to estimate the cosmic variance agree very well with those derived from other spectroscopic \citep{chapman03,chapman05} and photometric \citep{pope05,aretxaga03,aretxaga07,wardlow11} redshift estimates, which find $\bar{z} = 2.2-2.5$ and $\Delta z = 1.1-1.8$. The largest uncertainty in this estimate comes from the assumed stellar mass distribution of SMGs. Stellar mass estimates for SMGs are highly uncertain, as they depend strongly on the choice of the stellar synthesis model, star-formation history, and initial mass function (IMF) -- all of which are not well understood \citep[see discussion in][for example]{michalowski11}. The stellar mass distribution of AzTEC/GOODS SMGs in \citet[][see their figure~8]{yun11} peaks at $M_\star \sim 10^{11.3}\,\mathrm{M}_\odot$ and is broadly consistent with the mean stellar masses of SMGs estimated in other works, which range from $\sim10^{10.8}$ to $10^{11.8}\,\mathrm{M}_\odot$ \citep{dye08,michalowski10, hainline11, wardlow11}. However, the distribution in $M_\star$ from all of these works is found to be quite broad, especially compared to the stellar mass bins used in \citet{moster11} for computing the galaxy bias; this is why we opt to use a lower stellar mass limit of $M_\star > 10^{10}\,\mathrm{M}_\odot$, which includes 75\,per cent of the \citet{yun11} sample. Increasing the lower limit on the stellar mass would increase the expected cosmic variance for these surveys, as more massive galaxies are more strongly clustered. If we instead assume $M_\star > 10^{10.5}\,\mathrm{M}_\odot$ \citep[as motivated to some extent by results in][see their figure~3]{michalowski11}, we would derive a root cosmic variance of 8.8--17.5\,per cent for the individual AzTEC surveys and 5.3\,per cent for the combined fields.

\subsection{Systematic uncertainty from flux calibration}
\label{ssec:calerror}

We must also consider a systematic uncertainty on the derived source counts arising from uncertainty in the absolute flux calibration of our data. For AzTEC data, we determine flux conversion factors to convert the raw detector signals to flux density units based on several calibration observations of planets taken over a wide range of atmospheric conditions \citep[see][for details]{wilson08}. While the random calibration error of an individual observation is $\sim10$\,per cent \citep{scott10}, co-added AzTEC maps are each built from 46 to 325 observations, and the error on our measured source flux densities integrates down to 0.3\,per cent when all observations, and all fields, are considered. However, for all of these data we use the same flux calibrators, Uranus and Neptune, which have an absolute uncertainty on their flux densities of $\sigma_{\mathrm{cal}} = 5$\,per cent \citep{griffin93}. This is a systematic uncertainty in the flux scale of our maps that is completely correlated among all AzTEC data and therefore propagates into the source counts.

As with cosmic variance, we incorporate this calibration uncertainty into the bootstrap sampling method. For each of the 20,000 iterations, we generate a random number drawn from a Gaussian distribution with a mean of zero and a standard deviation of $\sigma_{\mathrm{cal}}$. We then modify the PFD of {\it every} source assuming that the observed flux {\it and} noise change by this fractional amount, consistent with a systematic change in our flux calibration. We use the same method of sampling the PFDs as described in Section~\ref{ssec:bootstrap}, creating 20,000 realisations of the counts, from which we compute the mean, 68.3\,per cent confidence intervals, and the covariance matrix, as before.

The correlation matrices for the counts, when including this systematic calibration uncertainty, are shown in Tables~\ref{tab:alldcorr} and \ref{tab:allicorr}. The standard deviation on the differential counts increases by $\le 5$\,per cent in all flux bins. We find that the bin-to-bin correlations on the differential counts do not increase substantially; this is because the perturbations to the PFDs of the sources that account for the absolute calibration uncertainty are small compared to the intrinsic width of the PFDs owing to the low signal-to-noise of the detections. It is this latter feature that gives rise to the strong correlations seen among the bins {\it before} including any systematic uncertainties.

The differential source counts and 68.3\,per cent confidence intervals, when systematic uncertainties from both cosmic variance and flux calibration are included, are shown in Fig.~\ref{fig:dnc_models}. Tables~\ref{tab:allc}, \ref{tab:alldcorr}, and \ref{tab:allicorr} list the differential and integrated source counts, 68.3\,per cent confidence intervals, and correlation matrices, with and without systematic uncertainties.

\section{Bright source counts from ACES and the SPT}
\label{sec:brightcounts}

Over the flux densities for which these AzTEC blank-field surveys are sensitive, the source counts are well described by a Schechter function (Equation~\ref{eqn:schechter}), declining exponentially with increasing flux density, since highly luminous galaxies are quite rare. Recent surveys at submm and mm wavelengths covering $\gtrsim100$\,deg$^2$ have been achieved from ground- and space-based observatories, including the Atacama Cosmology Telescope \citep[ACT;][]{marriage11}, the SPT \citep{vieira10}, and the {\it Herschel Space Observatory} \citep{eales10,oliver10}. These surveys are turning out large numbers of extremely bright, rare objects that are not associated with known nearby systems or strong radio sources, having SEDs consistent with high-redshift, dusty star-forming galaxies. It has been shown that a significant number of these extremely bright systems detected by {\it Herschel}, with 500\,$\mu$m flux densities $S_{500} \gtrsim 100$\,mJy, are strongly lensed by foreground galaxies or structure \citep{negrello10, conley11}, with magnification factors of $\mu \sim 10$. These lensed galaxies are believed to  contribute significantly to the source counts at flux densities greater than those probed by our comparatively small AzTEC surveys. \citet{vieira10} have estimated the counts at $S_{1400} > 10$\,mJy for 1.4\,mm sources detected by the SPT, excluding those that are associated with nearby galaxies discovered by the Infrared Astronomical Satellite ({\it IRAS}) and those with synchrotron-dominated (as opposed to dust-dominated) SEDs. These are shown alongside the AzTEC blank-field counts in Fig.~\ref{fig:dnc_models}, where we have scaled the SPT counts to 1.1\,mm assuming a spectral index of $\alpha=2.65$, which corresponds to the average spectral index between the observed flux densities at 1.1 and 1.4\,mm for a starburst galaxy at $z=3$ \citep{yun02}. The SPT counts, which are believed to be dominated by strongly lensed galaxies, diverge from the exponential fall-off that would be extrapolated from the AzTEC blank-field counts.

During the 2007-2008 observing seasons on ASTE, we observed 37 individual, relatively small ($\sim 300$ arcmin$^2$) fields centred on known over-dense regions as part of the AzTEC Cluster Environment Survey (ACES). This survey was designed to study the SMG population towards biased fields and includes regions surrounding clusters and proto-cluster candidates from $z =$\ 0.05 - 6, covering a total area of 3.1 deg$^2$. A full description of ACES and first results will be discussed in Zeballos et al. (2012).

The ACES fields were not included in our source counts estimate since we specifically wanted to avoid known biased regions. Still, each of the ACES maps includes a relatively large area, far from the central over-dense region that is not expected to be influenced by the cluster environment. We see several bright ($\ge 8$\,mJy) SMGs located far from the biased regions of these ACES maps;  from these, we estimate the bright counts at 1.1\,mm, which are poorly constrained by our combined blank-field surveys. Since our goal is to fill in the information on the {\it unbiased} source counts in the flux range that is not well sampled by the AzTEC blank-field counts or the SPT counts, we apply masks and various selection criterion to the ACES data.  We first exclude any SMG detected within 2\,arcmin of the centre of the cluster or proto-cluster core. At $z > 0.3$ (the minimum redshift of the ACES clusters where $S_{1100} \ge 8$\,mJy sources are found), 2\,arcmin corresponds to $\ge0.5$\,Mpc, which is larger than the expected core radii for massive clusters with total masses $10^{14-15.7}\,M_\odot$ \citep[e.g.][]{navarro95, lloyddavies00,kay01}. This mask will therefore exclude the major sources of potential biases to the counts, including those from strong lensing and the Sunyaev-Zeldovich effect, as well as most cluster members. Next, we follow a similar analysis as in \citet{vieira10} to exclude nearby galaxies and synchrotron-dominated sources from this sample. We exclude sources which have 2MASS $K$-band and/or {\it IRAS} counterparts located within the AzTEC beam to eliminate low-redshift galaxies. We then check for sources associated with bright radio objects using the Australia Telescope Compact Array (ATCA) 20-GHz survey \citep{murphy10}, the Sydney University Molonglo Sky Survey (SUMSS) at 843 MHz \citep{bock99}, and a unified catalogue of radio sources, which combines information from FIRST, NVSS, WENSS, GB6, and SDSS \citep{kimball08}. None of the $\ge8$\,mJy ACES sources have radio associations, implying that their SEDs are likely dominated by dust emission. After culling our sample using the above selection criterion, we have a total of 35 $S_{1100} \ge 8$\,mJy sources from these ACES fields.

The ACES bright source counts are shown in Fig.~\ref{fig:dnc_models} and are listed in Table~\ref{tab:allc}. Given small sample statistics, the errors on the bright source counts derived from the ACES fields are large; however, we see tentative evidence for a divergence of the counts from an exponential fall-off at $S_{1100} \gtrsim 13$\,mJy, and a smooth connection between the AzTEC blank-field counts at $S_{1100} < 12$\,mJy and the SPT counts at $S_{1100} \gtrsim 20$\,mJy. This upturn may be highlighting the regime at which the source counts become dominated by gravitational lensing effects. On the other hand, it is possible that the 2\,arcmin mask is not large enough to remove all cluster members, and the upturn in the counts at $S_{1100} \gtrsim 13$\,mJy is merely reflecting an over-density of SMGs in the outer regions of the clusters. The source counts at $S_{1100} = 1-10$\,mJy derived from the ACES fields, however, do not support this; Zeballos et al. (2012) find that the ACES source counts are completely consistent with our results from blank-fields over these flux densities when the inner regions (1.5\,arcmin radii) are masked. Since a real over-density would be flux-independent, the lack of an excess in the ACES counts at $S_{1100} < 10$\,mJy suggests that the upturn at $S_{1100} \gtrsim 13$\,mJy is not due to cluster-member contamination.

\section{Comparison with galaxy evolution models}
\label{sec:galevol}

With these combined AzTEC surveys, we have put the strongest constraints to date on the blank-field 1.1\,mm source counts from $S_{1100} = 1$ to 12\,mJy. These counts provide important information for modelling the formation and evolution of galaxies. A detailed analysis of how these results fit into our current understanding of galaxy evolution is beyond the scope of this paper. However, we can compare our observed 1.1\,mm counts to existing predictions from evolutionary models from the literature, many of which have used constraints from 1.1\,mm source counts in the past from smaller and/or shallower surveys, in order to motivate modifications to existing models in light of these new constraints. We provide only a qualitative comparison here, since common statistical tests (e.g. the Pearson $\chi^2$ test or the Kolmogorov-Smirnov test) do not apply given the strong correlations between these binned data.

\begin{figure*}
\begin{center}
\includegraphics[width=13.5cm]{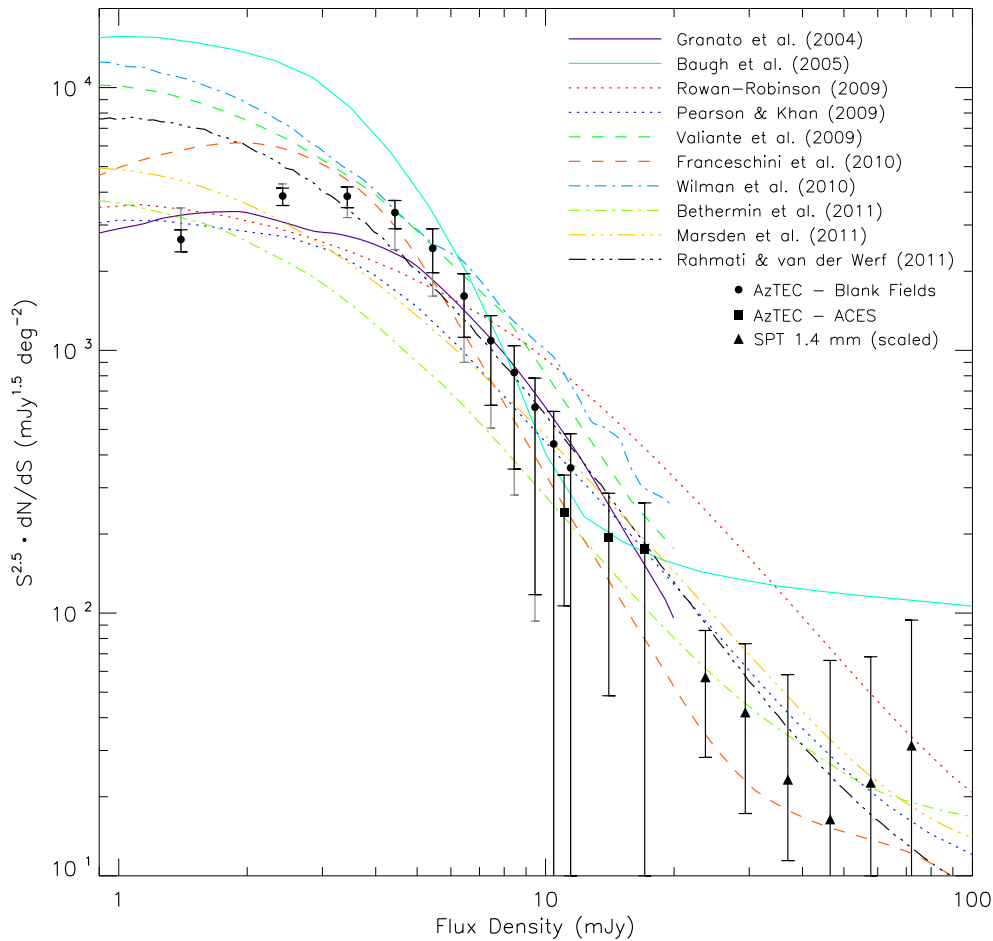}
\caption{Comparison of  the observed, Euclidean-normalised differential source counts from AzTEC surveys and predictions from galaxy evolution models. The differential counts derived from the combined six AzTEC blank fields are shown as circles (same as Fig.~\ref{fig:dnc_all}), where the black error bars show the 68.3\,per cent confidence interval, including systematic uncertainties from cosmic variance (Section~\ref{ssec:cosmicvar}) and flux calibration (Section~\ref{ssec:calerror}). The one-sided extended error bars shown in grey encompass the 68.3\,per cent confidence intervals including corrections to the measured counts due to bias from confusion effects, as discussed in Section~\ref{sec:galevol}. The squares show the bright source counts derived from ACES fields, as described in Section~\ref{sec:brightcounts}. The triangles show the 1.4\,mm source counts from the SPT survey \citep{vieira10}, excluding nearby {\it IRAS} galaxies and sources whose SEDs are dominated by synchrotron emission. The SPT data have been scaled to 1.1\,mm assuming a spectral index of $\alpha = 2.65$. The curves correspond to predictions from various semi-analytical and phenomenological models taken from the literature, as listed in the legend.}
\label{fig:dnc_models}
\end{center}
\end{figure*}

Ten such models are shown in Fig.~\ref{fig:dnc_models}. These include predictions from the semi-analytical models of \citet{granato04}, \citet{baugh05}, and \citet{wilman10}, and from the phenomenological models of \citet{rowanrobinson09}, \citet{pearson09}, \citet{valiante09}, \citet{franceschini10}, \citet{bethermin11}, \citet{marsden11}, and \citet{rahmati11}. All of these models assume the standard $\Lambda$CDM cosmology, but with slightly different parameters ranging from $\Lambda = 0.7$ to 0.734, $\Omega_{\mathrm{m}} = 0.266$ to 0.3, and $H_0 = 70$ to 75\,km\,s$^{-1}$\,Mpc$^{-1}$; we have not scaled the model predictions to a uniform cosmology since the differences should be minor. The models show a great deal of dispersion among themselves, differing by as much as a factor of four at any given flux density. With the exception of \citet{baugh05} and \citet{bethermin11}, all of the models appear to be largely consistent (within $\sim3\sigma$) with the observed source counts at $S_{1100} \gtrsim 4$\,mJy, but many are significantly discrepant at lower flux densities.

With the exception of \citet{bethermin11}, these models did not include the effects of strong lensing on the counts (which is expected to become important around $S_{1100} \gtrsim 15$\,mJy), and did not use the bright counts measured by the SPT as constraints. The general agreement between these models and the ACES and SPT counts is therefore somewhat fortuitous. However, this should not be over-interpreted, since the errors on the ACES and SPT counts are quite large, and the scaling used to convert the 1.4\,mm counts from the SPT survey to 1.1\,mm is uncertain. We thus limit our discussion in this section to the comparison of these models to the AzTEC blank-field counts.

Considering first the semi-analytical models, the model that best fits the observed 1.1\,mm source counts at all flux densities is that of \citet{granato04}, which considers the evolution of gas within dark matter halos as driven by gravity, radiative cooling, and feedback from supernovae and AGN. Alternatively, \citet{baugh05} modelled the 850\,$\mu$m counts available at the time within the context of hierarchical assembly, where bursts of star formation are triggered solely by galaxy mergers; their model requires a top-heavy IMF in order to explain the number density of SMGs. Although this model over-predicts the 1.1\,mm counts by $>3\sigma$ at $S_{1100} <5$\,mJy, minor changes to the model could bring it more in line with our measurements (e.g., changing the dust emissivity).

The semi-analytical model of \citet{wilman10} is an extension of their previous work to simulate the extragalactic radio continuum sky, including AGN and star-forming galaxies, within the framework of their large-scale clustering. Unlike \citet{granato04} and \citet{baugh05}, who use self-consistent radiative transfer calculations to describe the absorption and re-emission of starlight by dust, \citet{wilman10} use families of SED templates and FIR-radio relationships for star-forming and AGN-host galaxies to predict the IR-to-submm emission from their simulated radio galaxies, similar to what is done in phenomenological modelling. Their best-fit model accurately predicts the {\it Spitzer} 24--160\,$\mu$m counts, as well as the 850\,$\mu$m counts from SCUBA. However, we find that the model of \citet{wilman10} over-predicts the 1.1\,mm counts by $>3\sigma$ at $S_{1100} \lesssim 3$\,mJy. All three of the semi-analytical models considered here predate the large {\it Herschel} surveys, and it would be interesting to see how they fare when compared to the counts at shorter submm wavelengths.

We now compare the observed counts to predictions from phenomenological models. The only two that match the blank-field counts within $3\sigma$ at all flux densities are those of \citet[][with a redshift beyond which evolution is zero of $z_f = 4$]{rowanrobinson09} and \citet{pearson09}. Both of these models consider only four populations of galaxies, each represented by a single SED, and are therefore some of the simplest models presented here. However, they both used the published 1.1\,mm source counts from the AzTEC/GOODS-N field \citep{perera08} to constrain their models, so the agreement with the combined-field source counts is not surprising.

Like those of \citet{rowanrobinson09} and \citet{pearson09}, the phenomenological models of \citet{valiante09}, \citet{franceschini10}, and \citet{marsden11} used constraints from source counts ranging from 24 to $1100\,\mu$m, but did not use constraints on the 250--500\,$\mu$m counts from {\it Herschel} surveys -- though \citet{franceschini10} and \citet{marsden11} do consider counts at these wavelengths from BLAST data. It is interesting to note that all five of these models over-predict the {\it Herschel} $250-350\,\mu$m source counts \citep{oliver10,clements10,glenn10}. We find that the \citet{valiante09} and \citet{franceschini10} models provide a decent fit (within $3\sigma$) to the 1.1\,mm source counts in all but the two lowest flux bins. \citet{valiante09} did not use the source counts at 1.1\,mm to constrain their models, which at the time of their publication were limited to data from relatively small and/or shallow fields \citep{laurent05,bertoldi07,perera08}. In comparison, \citet{franceschini10} did use constraints on the 1.1\,mm counts from the AzTEC/JCMT survey of COSMOS \citep{austermann09}, and their model is in somewhat better agreement with our combined-field counts. However, compared to the other AzTEC fields shown in Fig.~\ref{fig:dnc_all}, the AzTEC/JCMT survey targeted a smaller and considerably over-dense region within COSMOS, so it is not too surprising that the model from \citet{franceschini10} now over-predicts the counts at low flux densities. \citet{marsden11} used the counts determined from the AzTEC/SHADES fields to fit their model, but it predates the correction to the AzTEC transfer function \citep{downes11}; it now therefore under-predicts the combined-field counts at flux densities $3\,\mathrm{mJy} < S_{1100} < 5$\,mJy, and over-predicts the counts in the lowest flux bin, by $>3\sigma$. However, this model is in better agreement with the counts at $S_{1100} < 3$\,mJy than those of \citet{valiante09} and \citet{franceschini10}.

We consider two phenomenological models that include recent constraints on the 250--500\,$\mu$m counts from {\it Herschel}. The \citet{bethermin11} model under-predicts the 1.1\,mm source counts by $>3\sigma$ from $2\,\mathrm{mJy} < S_{1100} < 6$\,mJy. Those authors used the 1.1\,mm counts from the AzTEC surveys of GOODS-S \citep{scott10} and SHADES \citep{austermann10} to constrain their model; however, the published counts from those surveys were systematically low due to the error in our transfer function estimate. On the other hand, \citet{rahmati11} did not use the 1.1\,mm counts from published AzTEC surveys to constrain their model; however, they did compare their best-fit model to these data and showed that they are largely consistent, though their model over-predicts the 1.1\,mm counts at $S_{1100} \lesssim3$\,mJy.

As demonstrated in Section~\ref{ssec:confusion}, we expect the extracted 1.1\,mm counts to be moderately biased due to confusion effects. Given the small statistical errors on the measured counts from these surveys, these biases should be taken into account when assessing the agreement between these data and various models. In Fig.~\ref{fig:dnc_models}, we include one-sided extended error bars (in grey) to account for the bias in our measurements due to confusion effects. These error bars encompass the 68.3\,per cent confidence interval on the {\it corrected} counts, where we have used the results from the simulations in Section~\ref{ssec:confusion} to estimate correction factors to the measured counts. While this is only an approximate correction for the effects of confusion, it is sufficient for our purpose of qualitatively comparing our measurements to predictions from galaxy evolution models. As can be seen in Fig.~\ref{fig:dnc_models}, extending the uncertainty on the counts naturally brings the data and models into somewhat better agreement; however, the discrepancies at $S_{1100} \gtrsim 4$\,mJy remain large, and the main conclusions from the comparisons to various evolution models above are unchanged.

\section{Discussion}
\label{sec:disc}

Given that galaxy evolution models using semi-analytical methods provide some insight into the physical processes occurring within galaxies (albeit with several simplistic assumptions), it is interesting to compare these predictions with our measured counts. The best-fit semi-analytical model to the combined 1.1\,mm counts is that of \citet{granato04}. That model was able to reproduce the 850\,$\mu$m source counts and their redshift distribution measured at that time, as well as the $K$-band luminosity function of massive spheroids at $z=1.5$. A key feature of their model is that feedback from SN is more effective in slowing down the rate of star formation in shallower potential wells, so that star formation progresses more rapidly within the most massive halos. This scenario seems to be consistent with recent evidence in favour of ``downsizing" of SMGs \citep[e.g][]{dye08}, where, contrary to the hierarchical collapse of dark matter, star formation in the early Universe predominately takes place within the most massive systems and progresses to lower mass systems at later times \citep[e.g.][]{cowie96,bundy06,franceschini06,mobasher09,magliocchetti11}. This idea is supported by the strong evolution of the luminosity function determined from phenomenological models, which implies that the most luminous star-forming galaxies (with IR luminosities $L_{\mathrm{IR}} \gtrsim 10^{11}\,\mathrm{L}_\odot$) dominate at $z \gtrsim 1.5$, while normal galaxies dominate at lower redshifts. 

It is interesting to note that, of the phenomenological models discussed here, only that of \citet{franceschini10} is consistent with the turnover in the Euclidean-normalised counts at $S_{1100} \lesssim 2$\,mJy; however, like all of these models, it significantly over-predicts the counts at these faint flux densities, which have until now been only poorly constrained. This observed turnover in our data is statistically significant. If we fit our measured counts near the apparent peak at $S_{1100} = 2.4$ and 3.4\,mJy assuming no evolution (i.e. flat in Euclidean-normalised space), we find that the observed counts at $S_{1100} = 1.4$\,mJy fall short of this no-evolution model by $5\sigma$. The discrepancies between the predictions from these phenomenological models and the observed counts at faint flux densities may be highlighting our limited knowledge of both the SEDs and the density of low luminosity ($L_{\mathrm{FIR}} < 10^{10}\,\mathrm{L}_\odot$) galaxies in the local Universe and, in turn, the faint end of the local luminosity function \citep[see][and references therein]{chapin09b}. In their phenomenological modelling of the 70--1100\,$\mu$m detected populations, \citet{marsden11} find that the model which best fits the observed counts and redshift distributions over-predicts the CIRB at these wavelengths. They explore whether simple modifications to the SEDs of low luminosity galaxies -- to which the source counts are not sensitive, but which dominate the CIRB -- can improve the fit to the CIRB. They find that by assigning the warmest SEDs to $L_{\mathrm{FIR}} < 10^9\,\mathrm{L}_\odot$ galaxies, they can bring the CIRB prediction into agreement with the observed value. It is therefore possible that poor knowledge of the SEDs of local faint galaxies also limits how well phenomenological models can predict the faint end of the 1.1\,mm source counts. Alternatively, evolutionary models that consider two distinct populations evolving separately -- where low luminosity galaxies evolve less strongly and thus reduce the number of cold galaxies at high redshift \citep[e.g.][]{valiante09} -- can also successfully match both the observed source counts and the CIRB. If this is the case, the turn over at the faint end of the AzTEC 1.1\,mm counts may be providing important new information on the evolution of low luminosity systems. 

At the other extreme, accounting for the observed number density of bright SMGs poses a significant challenge for existing theoretical models \citep[e.g.][]{hayward11}. As discussed in Section~\ref{sec:brightcounts}, there is evidence for a bias in the observed source counts at $S_{1100} \gtrsim 15$\,mJy arising from background SMGs that are strongly lensed by foreground structure \citep{vieira10}. Although lensing is not expected to significantly bias the observed 1.1\,mm counts at lower flux densities, current modelling of this effect is necessarily simplistic and not well informed by observations \citep{negrello07,paciga09,lima10,bethermin11}. \citet{austermann10} and \citet{aretxaga11} have shown that $S_{1100} \gtrsim 5$\,mJy sources are spatially correlated with $z \lesssim 1.1$ optical/IR galaxies in COSMOS, suggesting that galaxy-galaxy  and galaxy-group lensing at moderate amplification levels may bias the observed counts high, even at more modest flux densities. \citet{wang11} find a spatial correlation between foreground optical/IR galaxies from the Sloan Digitized Sky Survey (SDSS) and {\it Spitzer}/IRAC with high-redshift SMGs detected by {\it Herschel} in the Lockman-SWIRE field; since the redshift distributions of these different populations do not overlap significantly, this is strong evidence that the correlation arises from gravitational lensing. We have repeated the analysis of \citet{wang11} for the individual AzTEC fields for which SDSS and/or IRAC data are available, and find that only the COSMOS field shows a tentative ($\sim2\sigma$) correlation with the low-redshift galaxy samples -- consistent with the findings in \citet{aretxaga11}. Since we are sampling much smaller fields than the {\it Herschel} Lockman-SWIRE, our statistical power is limited. Still, these results hint that lensing may have a significant effect on the observed source counts, even at moderate flux densities.

The observed counts can also be significantly biased if a large fraction of the SMGs detected as single point sources in low-resolution surveys, such as those carried out with AzTEC, LABOCA, and {\it Herschel}, are actually multiple systems blended by the large beam. High-resolution ($\sim2$\,arcsec) interferometric imaging with the Submillimeter Array (SMA) of an unbiased, flux-limited ($S_{1100} > 5.5$\,mJy) sample of 15 AzTEC-detected SMGs discovered in COSMOS showed that only two (13\,per cent) are resolved into two, physically unassociated galaxies \citep{younger07,younger09}. 
On the other hand, simulations designed to match the observed source counts suggest that the fraction of multiple, blended galaxies for SMGs detected in low-resolution observations can be even higher. \citet{wang11b} predict that $\sim 1/3$ of the $S_{850} > 5$\,mJy SCUBA sources are actually multiple galaxies blended by the beam. Similarly, using the same kind of simulations described in Section~\ref{ssec:confusion}, \citet{scott10} demonstrated that $\sim25$\,per cent of SMGs detected as single point sources in the confusion-limited AzTEC map of GOODS-S are likely to be two or more sources blended together. However, our simulations in Section~\ref{ssec:confusion} suggest that this has only a small effect on the measured source counts.

There are nevertheless potential biases to the observed source counts that we have not studied in our simulations: in particular, the effects of galaxy clustering. There are currently only weak constraints on the clustering strength of bright SMGs \citep[e.g.][]{blain04,weiss09,lindner11,williams11}, and the clustering strength of faint mm-selected sources below the confusion limit has not been measured. If these faint, but much more numerous galaxies are strongly clustered, the biases to the measured 1.1\,mm source counts could be much larger than those predicted by our simulations in Section~\ref{ssec:confusion}. Our lack of knowledge on the clustering properties of sub-mJy SMGs precludes a more rigorous study of the potential biases to the measured source counts from galaxy clustering. The fraction of blended SMGs detected in low-resolution surveys, and whether this leads to significant biases to the source counts, can be addressed in future observations with ALMA.

\section{Summary and conclusions}
\label{sec:conc}

We have combined previously published data from six blank-field surveys at 1.1\,mm taken with AzTEC, totalling 1.6\,deg$^2$ in area with rms depths of 0.4--1.7\,mJy\,beam$^{-1}$, in order to derive the strongest constraints to date on the 1.1\,mm extragalactic source counts from $S_{1100} = 1-12$\,mJy. We use the well tested bootstrap sampling method on the source catalogues to derive the counts, which allows for an accurate estimate of statistical errors, including correlations among the selected flux bins. Given the large total area sampled, we expect a systematic uncertainty in the counts arising from cosmic variance of only 3.9\,per cent. As discussed in Sections~\ref{ssec:cosmicvar} and \ref{ssec:calerror}, it is important to include systematic uncertainties from cosmic variance and flux calibration ($\sim5$\,per cent) into the total error on the observed source counts, since these must be considered when using the counts to constrain parameters in galaxy evolution modelling \citep[as done in][for example]{bethermin11}. We have included these systematic uncertainties in the total errors reported on our combined-field source counts by incorporating them directly into the bootstrap sampling method, and we list these in Tables~\ref{tab:allc}, \ref{tab:alldcorr}, and \ref{tab:allicorr}.

Comparing the observed 1.1\,mm source counts to predictions from several galaxy evolution models, we find that the agreement at flux densities $S_{1100} \gtrsim 4$\,mJy is generally good. Given that most of these models had been fit to bright source counts at 850\,$\mu$m and/or 1.1\,mm from previously published surveys, this agreement is expected. However, we find significant ($\gtrsim 3\sigma$) discrepancies between the combined-field 1.1\,mm counts and many of these models at $S_{1100} \lesssim 4$\,mJy. Similarly, with the exception of the most recent phenomenological models that include constraints from {\it Herschel} surveys, many of the models over-predict the counts at 250--500\,$\mu$m. The data presented here provide strong constraints -- highly complementary to those from {\it Herschel} surveys at shorter wavelengths -- that should be used in future modelling of the formation and evolution of IR-bright galaxies.

Of the semi-analytical models considered in this paper, the model of \citet{granato04} provides the best match to the 1.1\,mm source counts. This model is consistent with downsizing, in which the bulk of star formation activity progresses from more massive to less massive galaxies over time -- a scenario that is also supported by most phenomenological models that can describe the counts and redshift distribution of IR-bright sources. However, for the model of \citet{baugh05}, where the build-up of stellar systems is consistent with hierarchical formation, only minor modifications are required to improve its agreement with the observed source counts at 1.1\,mm. A better understanding of the physical processes of gas cooling and feedback are needed in order to provide information on which scenario bests describes the assembly of massive galaxies.

For the first time, we have been able to strongly constrain the 1.1\,mm counts at $S_{1100} < 3$\,mJy, and we measure a turnover in the Euclidean-normalised counts at $S_{1100} \lesssim 2$\,mJy which none of the evolutionary models considered here are able to reproduce. This either reflects our limited knowledge of the SEDs of low luminosity galaxies in the local Universe, or motivates modifications to the evolution of faint galaxies at high redshift. Wide area surveys at 60--500\,$\mu$m with {\it Herschel} \citep[e.g. H-ATLAS,][]{eales10} and future surveys at 450\,$\mu$m with SCUBA-2 and at 1.1\,mm with AzTEC on the Large Millimeter Telescope (LMT) will provide measurements of the SEDs of a large, unbiased sample of faint nearby galaxies, and in turn, allow for improved modelling of the evolution of these systems at high redshift.

While there is already considerable evidence that the counts at $S_{1100} \gtrsim 15$\,mJy are biased high by strong lensing effects, some groups have also demonstrated that galaxy-galaxy and galaxy-group lensing with moderate amplification may affect the observed counts at more modest flux densities ($S_{1100} \sim 5$\,mJy). Furthermore, galaxies detected as single point sources in low-resolution surveys, such as those taken with AzTEC, can also bias our measurements of the source counts if a significant fraction of these ``sources" are really two or more galaxies blended by the large beam. The effects of lensing and blending on the measured source counts are poorly understood, and may lead to inaccurate predictions from galaxy evolution models. High-resolution imaging of statistically significant samples of SMGs with ALMA will be possible in the near future, and these will greatly aid in quantifying the degree to which lensing and blended galaxies bias the observed source counts.

Compared to observations at shorter submm wavelengths from {\it Herschel}, surveys at 1.1\,mm sample (on average) galaxies at higher redshifts, as the negative $k$-correction for observations at longer wavelengths extends to $z \sim 10$. Indeed, a growing number of 1.1\,mm-bright SMGs are found to be at $z > 4$ \citep{daddi09b,daddi09a,coppin09,riechers10,capak11,smolcic11}. The source counts at 1.1\,mm presented here, and those derived from future surveys with AzTEC on the LMT, will therefore provide crucial information on the evolution of star-forming galaxies within the first $\sim2$\,Gyr after the Big Bang, where observations at shorter submm wavelengths provide few constraints.

\section*{Acknowledgements}

We thank Chris Pearson and Alberto Franceschini for providing us with predictions of the 1.1\,mm source counts from their galaxy evolution models. KSS is supported by the National Radio Astronomy Observatory, which is a facility of the National Science Foundation operated under cooperative agreement by Associated Universities, Inc. IA and DHH acknowledge support from CONACyT projects \#39953-F. BH is supported by a Research Fellowship for Young Scientists from the Japan Society of the Promotion of Science (JSPS). JSD acknowledges the support of the Royal Society via a Wolfson Research Merit award, and also the support of the European Research Council via the award of an Advanced Grant. This work is supported in part by grants \#0907952 and \#0838222 from the National Science Foundation. This work is supported in part by the MEXT Grant-in-Aid for Specially Promoted Research (\#20001003) and Scientific Research on Priority Areas (\#15071202). The ASTE project is driven by the Nobeyama Radio Observatory (NRO), a branch of the NAOJ, in collaboration with the University of Chile and Japanese institutes including the University of Tokyo, Nagoya University, Osaka Prefecture University, Ibaraki University and Hokkaido University. The James Clerk Maxwell Telescope is operated by the Joint Astronomy Centre on behalf of the Science and Technology Facilities Council of the United Kingdom, the Netherlands Organisation for Scientific Research, and the National Research Council of Canada.

\bibliographystyle{mn2e-alt}
\bibliography{comb_nc_bib}

\bsp

\label{lastpage}

\end{document}